\documentclass[twocolumn,pre]{revtex4-2}

\usepackage{amsmath,amssymb}
\usepackage{graphicx}
\usepackage{dcolumn}
\usepackage{bm}
\usepackage{xr}

\newcommand{\mass}{\underline{\underline{m}}}
\newcommand{\mobility}{\bm{\mu}}
\newcommand{\friction}{\bm{\zeta}}
\newcommand{\diffusivity}{\bm{D}}
\newcommand{\x}{\bm{x}}
\newcommand{\vel}{\bm{v}}
\newcommand{\de}{\mathrm{d}}

\newcommand{\Dbare}{D_\mathrm{bare}}
\newcommand{\kB}{k_\mathrm{B}}
\newcommand{\dt}{\de t}

\begin{document}

\title{Jensen bound for the entropy production rate in stochastic thermodynamics}

\author{Matthew P.\ Leighton}
\email{matthew\_leighton@sfu.ca}
\author{David A.\ Sivak}%
\email{dsivak@sfu.ca}
\affiliation{Department of Physics, Simon Fraser University, Burnaby, British Columbia, V5A 1S6, Canada.}%

\date{\today}

\begin{abstract}
Bounding and estimating entropy production has long been an important goal of nonequilibrium thermodynamics. We recently derived a lower bound on the total and subsystem entropy production rates of continuous stochastic systems. This `Jensen bound' has led to fundamental limits on the performance of collective transport systems and permitted thermodynamic inference of free-energy transduction between components of bipartite molecular machines. Our original derivation relied on a number of assumptions, which restricted the bound's regime of applicability. Here we derive the Jensen bound far more generally for multipartite overdamped Langevin dynamics. We then consider several extensions, allowing for position-dependent diffusion coefficients, underdamped dynamics, and non-multipartite overdamped dynamics. Our results extend the Jensen bound to a far broader class of systems.
\end{abstract}

\maketitle


\section{Introduction}
Bounds on entropy production are useful tools in thermodynamics, placing constraints on the space of physically realizable systems and processes. For an example, we need look no further than the second law itself. The second law was famously used by Carnot to bound the efficiency of heat engines~\cite{carnot1824reflections}, leading to improved steam engine designs with greater efficiencies. Entropy-production bounds likewise serve as tools for thermodynamic inference~\cite{seifert2019stochastic}, where observations of a system's dynamical behavior are used to constrain hidden thermodynamic details.

The last decade has seen a flurry of entropy-production bounds derived through the framework of stochastic thermodynamics~\cite{seifert2012stochastic}, the most well-known being the thermodynamic uncertainty relation (TUR). First conjectured in Ref.~\cite{barato2015thermodynamic} and later proven more generally in Ref.~\cite{gingrich2016dissipation}, the TUR encompasses a family of relations bounding entropy production using fluctuations of observable currents (Ref.~\cite{horowitz2020thermodynamic} reviews recent developments). Important applications include bounding and inferring the efficiency of motor proteins from experimental data~\cite{pietzonka2016universal}. Other stochastic-thermodynamic entropy-production bounds include geometrical bounds based on the Wasserstein distance~\cite{nakazato2021geometrical} and various thermodynamic speed limits~\cite{shiraishi2018speed,falasco2020dissipation,van2020unified}.

In recent work~\cite{leighton2022dynamic}, we introduced a new lower bound on entropy production rates for multi-component stochastic systems which we call the Jensen bound. The formalism constrains both total and subsystem-specific entropy production rates using only bare friction or diffusion coefficients and mean rates of change of coordinates, all experimentally accessible quantities. Ref.~\cite{leighton2022dynamic} leveraged this framework to derive a collection of bounds and Pareto frontiers constraining the performance of collective motor-driven transport systems within cells. Ref.~\cite{leighton2023inferring} showcased different applications of the Jensen bound, using it to infer from limited experimental data the internal free-energy transduction and thus subsystem efficiencies in bipartite molecular machines.

The Jensen bound, as derived in Ref.~\cite{leighton2022dynamic}, relies on a handful of key assumptions. The system is assumed to be isothermal, without time-dependent external control (i.e., autonomous), and at a nonequilibrium steady state. Further the system is assumed to obey multipartite overdamped Langevin dynamics, and all degrees of freedom are assumed to have equal mean rates of change. While these assumptions already encompass a broad class of stochastic systems, they nonetheless limit the applicability of the Jensen bound.

In this paper we significantly extend the Jensen-bound framework, providing a general derivation for continuous stochastic systems with multiple degrees of freedom. In particular, we focus on four main objectives. We first derive the Jensen bound in full generality for multipartite overdamped Langevin dynamics, allowing for deviations from Ref.~\cite{leighton2022dynamic} including external control, different temperatures, and different mean coordinate rates of change; we also relax the steady-state assumption. We then establish three additional extensions to the Jensen bound: inhomogeneous diffusion coefficients, underdamped dynamics, and non-multipartite overdamped dynamics. These derivations place the Jensen bound on more general theoretical ground, and significantly broaden its range of applicability. We also explore the relationship between the Jensen bound and the TUR, showing explicitly how for some classes of systems the Jensen bound can be derived from short-time limits of different TUR formulations.

\section{Multipartite overdamped Langevin Dynamics}
Consider an overdamped stochastic system with $N$ degrees of freedom, denoted $\x=\{x_i\}_{i=1}^N$, each in contact with a heat bath at temperature $T_i$. The system is characterized by a mobility tensor $\mobility$, which is related to the hydrodynamic friction tensor $\friction$ by $\mobility = \friction^{-1}$~\cite{cichocki1994friction,cichocki2000friction}, as well as a diffusion tensor $\diffusivity$. For multipartite systems the mobility, friction, and diffusion tensors are all diagonal, so that each coordinate is subject to independent thermal noise~\cite{horowitz2015multipartite}. We assume their elements to be independent of time and position (this assumption will be relaxed in the next section). $D_i$, $\zeta_i$, and $\mu_i$ denote the coefficients of the $i$th subsystem.
 
The system dynamics depend on the forces that influence it; these forces may either result from a conservative potential $V(\x,t)$, or be nonconservative forces $\bm{f}_\mathrm{nc}(\x,t)$ which do not arise from a potential. Both conservative and nonconservative forces may in general be time-dependent due to external control. The total force vector acting on the system is then $\bm{f}(\x,t) = \bm{f}_\mathrm{nc}(\x,t) - \nabla V(\x,t)$.

The system evolves dynamically according to a multidimensional overdamped Langevin equation,
\begin{equation}
\dot{\bm{x}} = \mobility \bm{f}(\x,t) + \bm{\eta}(t),
\end{equation}
where the random fluctuations $\bm{\eta}(t)$ have zero mean and correlations satisfying
\begin{equation}
\left\langle \bm{\eta}(t)\bm{\eta}^\top(t')\right\rangle = 2 \diffusivity \, \delta(t-t').
\end{equation}
We assume the fluctuation-dissipation theorem, so that elements of the diffusion and mobility tensors are related by~\cite{maes2008steady,ito2020stochastic}
\begin{equation}\label{eq:fdt}
D_{ii} = \mu_{ii} \, \kB T_i,
\end{equation}
for Boltzmann's constant $\kB$.

These dynamics can equivalently be described at the level of the probability distribution $p(\x,t)$ by a Fokker-Planck equation (FPE):
\begin{equation}\label{General_FPE}
\frac{\partial}{\partial t} p(\x,t) = -\nabla\cdot \bm{J}(\x,t),
\end{equation}
with probability flux vector
\begin{equation}\label{fluxdef}
\bm{J}(\x,t) \equiv \mobility \, \bm{f}(\x,t) \, p(\x,t) - \diffusivity \, \nabla p(\x,t).
\end{equation}

Following Ref.~\cite{horowitz2015multipartite}, the entropy production rate (scaled by $\kB$) is
\begin{equation}
\dot{\Sigma} = \sum_{i=1}^N D_i^{-1}\left\langle \left[\frac{J_i(\x,t)}{p(\x,t)}\right]^2\right\rangle,
\end{equation}
which can then be split into contributions from each subsystem, so that the entropy production rate of the $i$th subsystem is
\begin{equation}\label{eq:subsystemEPR}
\dot{\Sigma}_i =  D_i^{-1}\left\langle \left[\frac{J_i(\x,t)}{p(\x,t)}\right]^2\right\rangle.
\end{equation}
Here and elsewhere unless otherwise specified, angle brackets denote ensemble averages performed over the nonequilibrium probability distribution $p(\x,t)$ at a given time $t$. Due to the multipartite assumption, the total entropy production rate is simply the sum of all subsystem entropy production rates~\cite{horowitz2015multipartite},
\begin{equation}
\dot{\Sigma} = \sum_{i=1}^N \dot{\Sigma}_i.
\end{equation}

Given the functional form of Eq.~\eqref{eq:subsystemEPR}, Jensen's inequality~\cite{cover1999elements} dictates that
\begin{equation}
\left\langle \left[\frac{J_
i(\bm{x},t)}{p(\bm{x},t)}\right]^2\right\rangle\geq \left\langle \frac{J_i(\bm{x},t)}{p(\bm{x},t)}\right\rangle^2.
\end{equation}
On the right hand side we identify the mean rate of change $\langle \dot{x}_i\rangle$ of the $i$th coordinate, which is equal to the ensemble-averaged probability current~\cite{seifert2012stochastic}: 
\begin{equation}\label{eq:vAv}
\langle \dot{x}_i\rangle = \left\langle \frac{J_{i}(\bm{x},t)}{p(\bm{x},t)}\right\rangle.
\end{equation}

Two inequalities follow from this, respectively constraining the partial and total entropy production rates:
\begin{subequations}\label{entropy_inequalities}
\begin{align}
\dot{\Sigma}_i & \geq D_i^{-1}\langle \dot{x}_i\rangle^2,\label{localjensenbound}\\
\dot{\Sigma} & \geq \sum_{i=1}^ND_i^{-1}\langle \dot{x}_i\rangle^2.\label{globaljensenbound}
\end{align}
\end{subequations}
We call Eq.~\eqref{localjensenbound} the \textit{subsystem Jensen bound} and Eq.~\eqref{globaljensenbound} the \textit{total Jensen bound}. These can equivalently be written in terms of friction coefficients rather than diffusion coefficients.

When all subsystems have the same mean velocity $\langle v\rangle$, as is the case in the collective transport systems studied in Ref.~\cite{leighton2022dynamic}, Eq.~\eqref{globaljensenbound} simplifies further to
\begin{equation}\label{meanvjensenbound}
\dot{\Sigma} \geq \Dbare^{-1} \langle v\rangle^2.
\end{equation}
Here $\Dbare$ is the ``bare collective diffusivity'', the inverse of the total friction coefficient from summing the individual friction coefficients (inversely proportional to bare diffusivities) of each subsystem:
\begin{equation}
\Dbare \equiv \left(\sum_{i=1}^N D_i^{-1}\right)^{-1}.
\end{equation}
Physically, $\Dbare$ is the effective diffusivity under a potential that only depends on relative subsystem positions. 

The derivation in this section followed that in Ref.~\cite{leighton2022dynamic}, concluding with  Eqs.~\eqref{localjensenbound} and \eqref{globaljensenbound}. But here we have generalized beyond the restriction in \cite{leighton2022dynamic} to steady state, and allowed for time-dependent forces, different subsystem temperatures, and different mean coordinate rates of change. In the following sections, we further generalize the Jensen bound in three different directions to allow for position-dependent diffusion coefficients, underdamped dynamics, and non-multipartite dynamics.

\section{Position-dependent diffusion coefficients}
Suppose now that the system has position-dependent diffusion coefficients, such that $D_i = D_i(\x)$. By the fluctuation-dissipation theorem~\eqref{eq:fdt}, the mobility and friction coefficients thus also depend on position. The entropy production rate for the $i$th subsystem is then~\cite{van2010three}
\begin{equation}\label{positiondependentDEPR}
\dot{\Sigma}_i =  \left\langle D_i(\x)^{-1} \left[\frac{J_i(\x,t)}{p(\x,t)}\right]^2\right\rangle.
\end{equation}
To bound this quantity, we use the Cauchy-Schwarz inequality, which applied to the covariance of two random variables $X$ and $Y$ gives~\cite{mukhopadhyay2020probability}
\begin{equation}
\left\langle X Y\right\rangle^2\leq\left\langle X^2\right\rangle\left\langle Y^2\right\rangle.
\end{equation}
Specializing to $X = J_i(\x,t)/[\sqrt{D_i(\x)}p(\x,t)]$ and $Y=\sqrt{D_i(\x)}$ then gives
\begin{equation}
\left\langle \frac{J_i(\x,t)}{p(\x,t)}\right\rangle^2\leq\left\langle D_i(\x)^{-1} \left[\frac{J_i(\x,t)}{p(\x,t)}\right]^2\right\rangle\left\langle D_i(\x)\right\rangle.
\end{equation}
Identifying the entropy production rate $\dot{\Sigma}_i$~\eqref{positiondependentDEPR} on the right hand side and mean coordinate rate of change $\langle \dot{x}_i\rangle$~\eqref{eq:vAv} on the left hand side, dividing both sides by $\langle D_i(\x)\rangle$ then yields the subsystem Jensen bound
\begin{equation}\label{localjensenbound_posdep}
\dot{\Sigma}_i \geq \langle D_i(\x)\rangle^{-1}\langle \dot{x}_i\rangle^2,
\end{equation}
where the constant diffusion coefficient of Eq.~\eqref{localjensenbound} has been replaced by the steady-state average diffusion coefficient $\langle D_i(\x)\rangle$. Summing the subsystem Jensen bounds for all $N$ subsystems then yields the total Jensen bound:
\begin{equation}\label{globaljensenbound_posdep}
\dot{\Sigma} \geq \sum_{i=1}^N \langle D_i(\x)\rangle^{-1}\langle \dot{x}_i\rangle^2.
\end{equation}
While these expressions require knowledge of the nonequilibrium probability distribution to evaluate the quantities $\left\langle D_i(\x)\right\rangle$, the entropy production rates can be further lower bounded using the maximum values of the diffusion coefficients,
\begin{equation}
D_i^{\mathrm{max}} \equiv \mathrm{max}\left\{ D_i(\x)\,:\, \x\in \mathrm{dom}(\x)\right\}.
\end{equation}
Here $\mathrm{dom}(\x)$ is the domain of $\x$ over which the function $D_i(\x)$ is defined. Since $D_i^\mathrm{max}\geq D_i(\x)$ for all $\x\in\mathrm{dom}(\x)$, we have
\begin{subequations}
\begin{align}
\dot{\Sigma}_i & \geq \left(D_i^\mathrm{max}\right)^{-1}\langle \dot{x}_i\rangle^2,\\
\dot{\Sigma} & \geq \sum_{i=1}^N \left(D_i^\mathrm{max}\right)^{-1}\langle \dot{x}_i\rangle^2.
\end{align}
\end{subequations}

\section{Alternative derivation from the short-time TUR}\label{shorttimeturderivation}
Here we show that the subsystem Jensen bounds~\eqref{localjensenbound} and \eqref{localjensenbound_posdep} can also be derived from the short-time thermodynamic uncertainty relation. In Ref.~\cite{otsubo2020estimating}, it was shown that for overdamped Langevin dynamics the entropy production rate of the $i$th subsystem is lower bounded by
\begin{equation}\label{eq:shorttimeTUR}
\dot{\Sigma}_i \geq \frac{2\left\langle j_{d_i}\right\rangle^2}{\mathrm{Var}\left(j_{d_i}\right)\dt},
\end{equation}
for any current
\begin{equation}
j_{d_i}\dt = d_i(\x,t)\circ \de x_i(t)
\end{equation}
in the limit as $\dt\to 0$. Setting $d_i(\x,t) = 1$ and computing the mean and variance of the short-time current following the methods of Ref.~\cite{otsubo2020estimating} gives $\left\langle j_{d_i}\right\rangle  = \left\langle \dot{x}_i\right\rangle$ and $\mathrm{Var}\left(j_{d_i}^\tau\right)\dt = 2\left\langle D_i\right\rangle$. Inserting these identities into the short-time TUR~\eqref{eq:shorttimeTUR} yields
\begin{equation}
\dot{\Sigma}_i \geq \langle D_i(\x)\rangle^{-1}\langle \dot{x}_i\rangle^2.
\end{equation}
This is identical to the subsystem Jensen bound for multipartite overdamped Langevin dynamics with position-dependent diffusion coefficients~\eqref{localjensenbound_posdep}, and simplifies to Eq.~\eqref{localjensenbound} for constant diffusion coefficients.

\section{Underdamped Langevin dynamics}
We now turn to underdamped Langevin dynamics, for which the equations of motion are~\cite{van2019uncertainty}
\begin{subequations}
\begin{align}
\dot{\x} & = \vel,\\
\mass\dot{\vel} & = -\friction\vel + \bm{f}(\x,t) + \bm{\xi}(t).
\end{align}
\end{subequations}
Here the random noise $\bm{\xi}(t)$ directly affects the velocity dynamics rather than position dynamics. $\bm{\xi}(t)$ is Gaussian and satisfies
\begin{subequations}
\begin{align}
\left\langle \bm{\xi}(t)\right\rangle & = 0,\\
\left\langle \xi_i(t)\xi_j(t')\right\rangle & = 2\kB T_i\,\zeta_{ij}\,\delta_{ij}  \, \delta(t-t').
\end{align}
\end{subequations}

The entropy production rate is~\cite{dechant2018entropic,van2019uncertainty}
\begin{equation}\label{underdampedEPR}
\begin{aligned}
\dot{\Sigma} & = \sum_{i=1}\underbrace{\left\langle \frac{m_i^2}{T_i\zeta_i}\left[\frac{J_{v_i}^\mathrm{irr}(\x,\vel,t)}{p(\x,\vel,t)}\right]^2\right\rangle}_{\dot{\Sigma}_i},
\end{aligned}
\end{equation}
for irreversible current
\begin{equation} 
J_{v_i}^\mathrm{irr}(\x,\vel,t) = \frac{1}{m_i}\left(-\zeta_iv_i - \frac{\kB T_i\zeta_i}{m_i}\frac{\partial}{\partial v_i}\right) p(\x,\vel,t).
\end{equation}
In this section, angle brackets denote ensemble averages over the joint probability distribution $p(\x,\bm{v},t)$ of positions and velocities at time $t$. 

Following Ref.~\cite{dechant2022bounds}, we write the velocity as $v_i = \nu_i(\x) + v_i - \nu_i(\x)$ for
\emph{local mean velocity}
\begin{equation}
\nu_i(\x) \equiv \int v_i \, p(v_i|\x) \, dv_i,
\end{equation}
and thereby decompose the entropy production rate into the sum of two non-negative terms:
\begin{align}
& \dot{\Sigma} = \sum_{i=1}^N D_i^{-1}\left\langle\nu_i(\x)^2\right\rangle \\
& + \sum_{i=1}^ND_i^{-1}\left\langle\left[ v_i-\nu_i(\x) +\frac{\kB T_i}{m_i}\frac{\partial}{\partial_{v_i}}\ln p(\bm{v},\x,t) \right]^2\right\rangle.\nonumber
\end{align}
Here we have used the fluctuation-dissipation relation~\eqref{eq:fdt} to rewrite friction coefficients in terms of diffusion coefficients.

Since both terms are non-negative, the first term is itself a lower bound for the entropy production. Applying Jensen's inequality to the first term, we arrive at subsystem and total Jensen bounds for underdamped Langevin dynamics:
\begin{subequations}\label{entropy_inequalities_underdamped}
\begin{align}
\dot{\Sigma}_i & \geq D_i^{-1}\langle v_i\rangle^2,\label{localjensenbound_underdamped}\\
\dot{\Sigma} & \geq \sum_{i=1}^ND_i^{-1}\langle v_i\rangle^2.\label{globaljensenbound_underdamped}
\end{align}
\end{subequations}

\section{Non-multipartite dynamics}
The derivations thus far have relied on the assumption of multipartite dynamics, namely that the friction ($\friction$), mobility ($\mobility$), and diffusion ($\diffusivity$) tensors are all diagonal. We now turn to the case where the multipartite assumption breaks down such that $\mobility$, $\friction$, and $\diffusivity$ may have non-zero off-diagonal elements. We restrict our attention to the isothermal case where all fluctuations arise from coupling to heat baths at temperature $T$, and to the case where all diffusion, friction, and mobility coefficients are constant. Crucially, we still assume the multidimensional fluctuation-dissipation theorem
\begin{equation}
\friction^{-1} = \mobility = \diffusivity /(\kB T).
\end{equation}
Without the multipartite assumption, defining the rate of entropy production becomes much less straightforward. While Ref.~\cite{maes2008steady} sketches a derivation of our eventual result, Eq.~\eqref{nonmultieprdef}, the next two subsections provide a full derivation for pedagogical purposes. Starting from the definition of total entropy production as a sum of changes in internal entropy (scaled by $\kB$) and heat dissipated to external baths~\cite{seifert2005entropy},
\begin{equation}
\dot{\Sigma} \equiv \dot{S} - \dot{Q}/\left(\kB T\right),
\end{equation}
we then proceed to evaluate the two terms $\dot{S}$ and $\dot{Q}$.

\subsection{Defining the heat flow}
Non-multipartite dynamics pose significant difficulties for identifying heat flows. In the previous sections we assumed one-to-one interactions between system components and heat baths. This is incompatible with non-multipartite dynamics, so a given system component can now be subjected to fluctuations from distinct sources. It is for this reason that to make headway we assume a single temperature $T$, since this allows us to ascribe all fluctuations to a single global heat bath. 

To define the heat flow, we start with the rate of change $\dot{E}$ of the energy which can be decomposed into a sum of two contributions that we identify as rates of work and heat:
\begin{subequations}
\begin{align}
\dot{E} & = \frac{d}{dt}\left\langle V(\x,t)\right\rangle\\
& = \int d\x \, p(\x,t)\frac{\partial}{\partial t}V(\x,t) + \int d\x \, V(\x,t)\frac{\partial}{\partial t}p(\x,t)\\
& = \underbrace{\int d\x \, p(\x,t)\frac{\partial}{\partial t}V(\x,t) + \left\langle \bm{f}_\mathrm{nc}(\x,t)^\top\circ \dot{\bm{x}}\right\rangle}_{\dot{W}} \\
& + \underbrace{\int d\x \, V(\x,t)\frac{\partial}{\partial t}p(\x,t) - \left\langle \bm{f}_\mathrm{nc}(\x,t)^\top\circ \dot{\bm{x}}\right\rangle}_{\dot{Q}}.
\end{align}
\end{subequations}
Inserting the Fokker-Planck equation~\eqref{General_FPE} and integrating the heat term by parts gives the standard definition of heat for multipartite dynamics:
\begin{equation}\label{eq:heatdef}
\dot{Q} = -\left\langle \bm{f}(\x,t)^\top \bm{J}(\x,t)/p(\x,t)\right\rangle.
\end{equation}

\subsection{Defining the entropy production rate}
We now compute the mean rate of change of system entropy, using the FPE~\eqref{General_FPE} following the approach of Ref.~\cite{seifert2005entropy}, as
\begin{subequations}
\begin{align}
\dot{S} & \equiv \frac{\de}{\de t} 
\big\langle - \ln p(\x,t)
\big\rangle \\
& =  \left\langle -\frac{\partial_tp(\x,t)}{p(\x,t)} - \frac{\nabla^\top p(\x,t)}{p(\x,t)} \circ\dot{\x}\right\rangle\\
& = \left\langle - \frac{\nabla^\top p(\x,t)}{p(\x,t)} \circ\dot{\x}\right\rangle.
\end{align}
\end{subequations}
Here $\circ$ denotes a Stratonovich product. In the third line the first term has been integrated out, since probability conservation imposes $\int \partial_tp(\x,t)d\x=0$.

This expression can be simplified further using the definition of the probability flux vector~\eqref{fluxdef}, rearranged to read
\begin{equation}
-\nabla^\top p(\x,t) = \left(\diffusivity^{-1}\right)^\top \bm{J}(\x,t)^\top - \frac{1}{\kB T}\bm{f}(\x,t)^\top p(\x,t),
\end{equation}
which yields
\begin{subequations}
\begin{align}
\dot{S} & = - \frac{1}{\kB T}\left\langle \bm{f}(\x,t)^\top \circ\dot{\x}\right\rangle + \left\langle \frac{\bm{J}(\x,t)^\top}{p(\x,t)} \left(\diffusivity^{-1}\right)^\top \circ \dot{\x}\right\rangle\\
& = -\frac{1}{\kB T} \left\langle  \bm{f}(\x,t)^\top \, \frac{\bm{J}(\x,t)}{p(\x,t)}\right\rangle \\
& \quad + \left\langle \frac{\bm{J}(\x,t)^\top \diffusivity^{-1} \, \bm{J}(\x,t)}{p(\x,t)^2}\right\rangle.\nonumber
\end{align}
\end{subequations}
In the last line we replaced the ensemble-averaged Stratonovich multiplication by $\dot{\x}$ with ensemble-averaged multiplication by the local mean velocity $\bm{J}(\x,t)/p(\x,t)$~\cite{seifert2012stochastic}, and replaced $\left(\diffusivity^{-1}\right)^\top$ with $\diffusivity^{-1}$ since the quadratic form in the second term is unchanged by taking the transpose of the matrix.

Finally, defining the total rate of entropy production as the sum of heat dissipated to the bath and system entropy change, 
\begin{subequations}\label{nonmultieprdef}
\begin{align}
\dot{\Sigma} & \equiv \dot{S} - \dot{Q}/\left(\kB T\right)\\
& = \left\langle \frac{\bm{J}(\x,t)^\top \diffusivity^{-1} \, \bm{J}(\x,t)}{p(\x,t)^2}\right\rangle.\label{eq:epreqnonmulti}
\end{align}
\end{subequations}
This agrees with the result derived in Ref.~\cite{maes2008steady} and the result reported in Ref.~\cite{ito2020stochastic}. Unlike in the case of multipartite dynamics discussed previously, for non-multipartite dynamics it is not generally possible to define non-negative entropy production rates at the subsystem level~\cite{chetrite2019information}.

\subsection{Jensen lower bound}
While more complicated than the multipartite case, the functional form for the entropy production rate obtained in Eq.~\eqref{nonmultieprdef} is still amenable to lower-bounding via Jensen's inequality. For the second law ($\dot{\Sigma}\geq 0$) to hold for any flux vector $\bm{J}$, the inverse of the diffusion matrix $\diffusivity^{-1}$ (or equivalently the friction matrix) must be positive semidefinite, a standard assumption for multi-dimensional friction matrices in stochastic dynamics~\cite{cichocki2000friction,ao2004potential,ito2020stochastic}. Note that the expression inside the angle brackets in Eq.~\eqref{eq:epreqnonmulti} is a quadratic form in the vector $\bm{J}(\x,t)/p(\x,t)$~\cite{boyd2004convex}. When $\diffusivity^{-1}$ is positive semidefinite, the quadratic form is a convex function of $\bm{J}(\x,t)/p(\x,t)$. Thus Jensen's inequality yields a general lower bound on the entropy production rate:
\begin{subequations}
\begin{align}
\dot{\Sigma} &  = \left\langle \frac{\bm{J}(\x,t)^\top \diffusivity^{-1} \, \bm{J}(\x,t)}{p(\x,t)^2}\right\rangle\\
& \geq \left\langle \frac{\bm{J}(\x,t)}{p(\x,t)}\right\rangle^\top \diffusivity^{-1} \left\langle \frac{\bm{J}(\x,t)}{p(\x,t)}\right\rangle\\
& = \left\langle\dot{\x}\right\rangle^\top\diffusivity^{-1}\left\langle\dot{\x}\right\rangle. \label{nonmultiJensen}
\end{align}
\end{subequations}
The result is a generalized Jensen bound depending only on the diffusion matrix and mean rates of change of the $N$ coordinates. This reduces to Eq.~\eqref{globaljensenbound} for multipartite dynamics when the diffusion matrix (or equivalently the friction matrix) is diagonal. 

When all system components have the same mean velocity, as is the case for collective transport systems~\cite{leighton2022performance,leighton2022dynamic}, Eq.~\eqref{nonmultiJensen} significantly simplifies to
\begin{equation}
 \dot{\Sigma} \geq \langle v\rangle^2\sum_{i,j}D_{ij}^{-1}.
\end{equation}
When the sum of all off-diagonal terms is non-negative, this can be further lower-bounded by the total Jensen bound for multipartite dynamics,
\begin{equation}
\dot{\Sigma} \geq \langle v\rangle^2\sum_{i=1}^N D_{ii}^{-1}, 
\end{equation}
thus recovering the results of Ref.~\cite{leighton2022dynamic}.

\subsection{Alternate derivation from the multidimensional TUR}\label{mturderivation}
It is also possible to derive Eq.~\eqref{nonmultiJensen} by taking a short-time limit of the multidimensional TUR. As derived in Ref.~\cite{dechant2018multidimensional} the multidimensional TUR bounds the total entropy production rate of a system obeying overdamped Langevin dynamics without requiring the multipartite assumption:
\begin{equation}
\Sigma = \int_{0}^{t_\mathrm{f}}\dot{\Sigma}\,\dt \geq 2\left\langle \bm{j}_{\bm{d}}\right\rangle^\top \bm{C}^{-1}\left\langle \bm{j}_{\bm{d}}\right\rangle.
\end{equation}
Here $\bm{j}_{\bm{d}}$ is a time-integrated current of the form
\begin{equation}
\bm{j}_{\bm{d}} = \int_0^{t_\mathrm{f}}\bm{d}(\x,t)^\top\circ \dot{x}\, \dt
\end{equation}
for any function $\bm{d(\x,t)}$, and $\bm{C}$ is the covariance matrix for the current $\bm{j}_{\bm{d}}$, defined as
\begin{equation}
C_{ik} = \left\langle j_{d_i} j_{d_k}\right\rangle - \left\langle j_{d_i}\right\rangle \left\langle j_{d_k}\right\rangle.
\end{equation}
In the limit ${t_\mathrm{f}} = \dt \to 0$, we recover $\Sigma = \dot{\Sigma} \,\dt$ and 
\begin{equation}
\bm{j}_{\bm{d}}\,\dt = {\bm{d}}(\x,t)^\top\circ \de\bm{x}(t).
\end{equation}
Setting $d_i(\x,t)=1$ for all $i$ yields $\left\langle \bm{j}_{\bm{d}}\right\rangle = \left\langle\dot{\bm{x}}\right\rangle$, and $C_{ik}\dt = 2  D_{ik}$. Thus we recover the Jensen bound for non-multipartite dynamics, Eq.~\eqref{nonmultiJensen}, so long as the diffusion coefficients are constant.

\section{Discussion}
In this paper we generalized the Jensen-bound framework introduced in Ref.~\cite{leighton2022dynamic}, deriving Jensen bounds on both subsystem and total entropy production rates for multipartite overdamped Langevin dynamics in much greater generality (Eqs.~\eqref{localjensenbound} and \eqref{globaljensenbound}). We also derived extensions in several directions, allowing for position-dependent diffusion coefficients (Eqs.~\eqref{localjensenbound_posdep} and \eqref{globaljensenbound_posdep}), multipartite underdamped dynamics (Eqs.~\eqref{localjensenbound_underdamped} and \eqref{globaljensenbound_underdamped}), and non-multipartite overdamped dynamics (Eq.~\eqref{nonmultiJensen}). These results significantly broaden the applicability of the Jensen bound to a far wider class of stochastic systems.

In all cases, the Jensen bound lower-bounds the entropy production rate of a continuous stochastic system in terms of only bare diffusion coefficients and mean rates of change of the system's degrees of freedom. These quantities can all be estimated from trajectory data: diffusion coefficients can be inferred using statistical methods~\cite{dechant2018estimating,frishman2020learning}, and mean rates of change can be computed directly by taking time- or ensemble-averages. In all of our results, the diffusion coefficients can be replaced by friction coefficients using the fluctuation-dissipation relation~\eqref{eq:fdt}.

While this article is focused on derivations, we also wish to highlight useful applications of our results. We first derived the Jensen bound as a tool to obtain fundamental limits on the performance of multi-component molecular machines like collections of transport motors, as quantified by metrics like velocity and efficiency~\cite{leighton2022dynamic}. Such bounds depend only on measurable quantities, and are independent of details like the potential-energy landscape coupling components together. In Ref.~\cite{leighton2023inferring}, we used subsystem Jensen bounds to infer internal free-energy transduction and subsystem efficiencies in bipartite molecular machines from experimental data, also independent of details of coupling between components. By extending the domain of the Jensen bound, we broaden the regime of applicability for these results. 

As a further application, we note that the Jensen bound can be used to estimate entropy production rates in systems where they cannot be calculated exactly. Explicitly calculating the entropy production rate of a stochastic system requires knowledge of the full nonequilibrium probability distribution along with all of the conservative and non-conservative forces acting on the system. These are all in general difficult to measure or compute for systems with many degrees of freedom. The Jensen bound, however, is much more straightforward to compute from data, provided the friction/diffusion coefficients are previously known or can be measured: mean coordinate rates of change can be computed from far less data than would be required to compute the full nonequilibrium probability distribution.

It is natural to ask how the Jensen bound relates to the thermodynamic uncertainty relation (TUR), of which there are both short-time~\cite{otsubo2020estimating} and long-time~\cite{gingrich2016dissipation} formulations. We showed in Sec.~\ref{shorttimeturderivation} and \ref{mturderivation} that the subsystem and total Jensen bounds for overdamped Langevin dynamics can be derived from short-time TURs, meaning that the Jensen bound will generally be equal to or looser than the short-time TUR for optimal choices of current. By contrast we previously compared the Jensen bound to the long-time TUR~\cite{gingrich2016dissipation} in Ref.~\cite{leighton2022dynamic}, finding for a numerically simulated model of collective motor-driven transport that there is no general hierarchy between the two bounds: either can be tighter, depending on the regime explored, with the long-time TUR tighter when current fluctuations are small, and the Jensen bound tighter otherwise. We also showed that the Jensen bound is always saturated for linear collective transport systems. While some of our main results can be derived from different short-time versions of the TUR, this work derives the Jensen bound for several broad classes of dynamics using a unified set of techniques centered around applications of Jensen's inequality. This unified derivation framework highlights the universality of the Jensen bound.

For applications to experimental data, we believe the main difference in the utilities of these different entropy production bounds arises from the different information required to compute them. The TUR and its multidimensional generalizations~\cite{dechant2018multidimensional,dechant2021improving,koyuk2022thermodynamic,tanogami2023universal} generally require measurements of variances and covariances of currents, but do not require detailed knowledge (\emph{e.g.}, friction/diffusion coefficients, conservative and nonconservative forces) regarding the equations of motion from which those dynamics arise. By contrast, the Jensen bound requires measurements of mean coordinate rates of change and knowledge of the friction/diffusion coefficients characterizing the system. While the TUR can make use of any currents in the system (including heat or energy currents), the Jensen bound only holds for coordinate rates of change. Ultimately the choice of which bounds to use should be made through considerations of what information and experimental data are available to the user.

In this article we restricted our attention to continuous dynamics. Discrete master-equation dynamics, the other main class of models for nanoscale stochastic processes, represent a formidable challenge for applying the Jensen bound. Since they do not feature bare friction/diffusion coefficients, or continuously varying coordinates whose mean rate of change can be measured, discrete dynamics are currently beyond the scope of the arguments we have presented. Nevertheless, many discrete models of  classical physical processes are really coarse-grained approximations of some underlying continuous dynamics, which would themselves obey the Jensen bound as derived in this article. We have also refrained from considering velocity-dependent forces (like those arising from magnetic fields), which are known to break the TUR in some cases~\cite{chun2019effect}, or non-thermal sources of fluctuations like active noise~\cite{Mizuno2007_Nonequilibrium,Gallet2009_Power}.

\emph{Acknowledgments}.---
We thank Jannik Ehrich (SFU Physics) for helpful feedback on the manuscript. This work was supported by a Natural Sciences and Engineering Research Council of Canada (NSERC) CGS  Doctoral fellowship (M.P.L.), an NSERC Discovery Grant and Discovery Accelerator Supplement (D.A.S.), and a Tier-II Canada Research Chair (D.A.S.).

\bibliography{main}

\end{document}